\documentclass[final,3p,sort&compress]{elsarticle}

\usepackage{amssymb}
\usepackage{amsmath}

\journal{Physica A}

\newcommand\beq{\begin{equation}}
\newcommand\eeq{\end{equation}}
\newcommand\beqa{\begin{eqnarray}}
\newcommand\eeqa{\end{eqnarray}}

\def\bal#1\eal{\begin{align}#1\end{align}}
\newcommand{\eff}{\text{eff}}

\begin{document}

\begin{frontmatter}



\title{A heuristic approach for the densest packing fraction of hard-sphere mixtures}


\author[label1]{Andr\'es Santos\corref{cor1}}
\ead{andres@unex.es}
\ead[url]{https://www.eweb.unex.es/eweb/fisteor/andres/Cvitae/}
\cortext[cor1]{Corresponding author}

\affiliation[label1]{organization={Departamento de F{\'{i}}sica, Universidad de Extremadura},
           city={Badajoz},
            postcode={E-06006},
            country={Spain}}

\author[label2]{Mariano L\'opez de Haro}
\ead{malopez@unam.mx}

\affiliation[label2]{organization={Instituto de Energ\'{i}as Renovables, Universidad Nacional Aut\'onoma de M\'exico},
            city={Temixco},
            postcode={62580},
            state={Morelos},
            country={Mexico}}

\begin{abstract}
In a previous work, a simple approach to derive the jamming packing fraction of a hard-sphere mixture from the knowledge of the random close-packing fraction of the monocomponent system was proposed. Now, an extension of that approach is applied to provide an approximate formula for the densest packing fraction of a given hard-sphere mixture in terms of the fcc close-packing fraction of a monocomponent hard-sphere system and of a single parameter encapsulating the dependence on the size ratios and the number of spheres in the unit cell. Comparison with recent results for such densest packing fraction of binary and ternary systems is performed and reasonable agreement is obtained.

\end{abstract}




\end{frontmatter}


\section{Introduction}

The search for simplicity and wide range of applicability of analytical results in physical problems has been a longstanding goal. This is of course not alien to sphere packing problems which, starting with Kepler's conjecture and despite many interesting developments (some very recent ones), still represent a considerable challenge \cite{AW08}.

In our previous works with hard-sphere (HS) systems, we have addressed the problem of mapping the equation of state  of an arbitrary fluid mixture of given size distribution and composition at a certain packing fraction onto the one of an effective one-component HS fluid \cite{SYH99,HYS08,S12c,S16,SYHO17,SYH20}. In this endeavor, we have been relatively successful in providing simple rules to, on the one hand and using only two well defined parameters, derive the equation of state of a mixture (either discrete or polydisperse) once the one of the monocomponent system is available; and, on the other hand, and using only one of the previous parameters (as specified below), to provide an estimate for the jamming packing fraction of the mixture from the known random close-packing fraction of the monocomponent system \cite{SYHO17,SYH20}. We refer to this as the \emph{surplus} approach, which will be sketched below, a detailed account of which for $d$-dimensional HS may be found in Chapter 3 of the book by one of us \cite{S16} and in Ref.~\cite{SYHO17}.

It is fair to acknowledge here also the work of the Princeton group on the densest binary sphere packings (DBSP) \cite{HJST11,HST12,HST13}, as well as the fairly recent one by Koshoji et al.\ \cite{KKFO21,KO21,KO22} on the same system and on densest ternary sphere packings (DTSP), which rely heavily on geometrical arguments and constructions. It is clear that increasing the number of components also increases the difficulty of such constructions and, therefore, although admittedly well beyond the range where the \emph{surplus} approach worked for fluid systems,  it is not unreasonable to ask if following a similar approach may shed some light on its usefulness also for the densest packings. The aim of this paper is to address such a question.

The organization of the paper is as follows. In Sec.~\ref{sec1}, and in order to make the paper self-contained, a rather brief account of the \emph{surplus} approach for fluid mixtures is presented, thus providing the necessary background material for the subsequent development. This is complemented by the extension of the previous ideas to derive an approximate formula to compute the densest packing fraction of any given mixture in terms of the close-packing fraction of the monocomponent system. Section \ref{sec2} contains the comparison of the results of our approximate (heuristic) formula with the recent ones derived with geometric arguments and its corresponding discussion. Finally the paper is closed in Sec.~\ref{sec3}, where we provide some concluding remarks.

\section{The \emph{surplus} approach}
\label{sec1}

We begin this section by presenting the main ideas behind the \emph{surplus} approach for fluid mixtures \cite{S16}.
The starting point is that the excess Helmholtz free energy per particle, $a_{\text{mixt}}^{\text{ex}}(\phi)$, and the compressibility factor, $Z_{\text{mixt}}(\phi)={p}/{\rho k_B T}$ (where $p$ is the pressure, $\rho$  is the number density,   $k_B$ is the Boltzmann constant, and $T$ is the absolute temperature), of a multicomponent (either discrete or polydisperse)  HS mixture  at a packing fraction $\phi\equiv\frac{\pi}{6}\rho \sum_i x_i \sigma_i^3$ (where  $x_i$ and $\sigma_i$ are the mole fraction and the diameter of spheres of species $i$, respectively)  may be constructed from the ones of the monocomponent HS fluid, $a^{\text{ex}}(\phi_\eff)$ and $Z(\phi_\eff)$, calculated at an \emph{effective} packing fraction $\phi_{\eff}$. Assuming  that the Helmholtz free energy per particle is truncatable ({i.e.}, it depends only on the first three moments of the size distribution) \cite{SWC01,S02}, and applying certain consistency conditions, one may derive the following relationships \cite{S16,SYHO17,SYH20},
\begin{subequations}
\label{HFEsp-Zsp}
\begin{equation}
\label{HFEsp}
\frac{a_{\text{mixt}}^{\text{ex}}(\phi)}{k_BT} + \ln (1-\phi)=\frac{\mu}{\lambda}\left[\frac{a^{\text{ex}}(\phi_\eff)}{k_BT} +  \ln (1-\phi_{\eff})\right],
\end{equation}
\begin{equation}
\label{Zsp}
 \phi\left[ Z_{\text{mixt}}(\phi)-\frac{1}{1-\phi}\right] =\mu \phi_{\eff}\left[ Z(\phi_{\eff})-\frac{1}{1-\phi_{\eff}}\right],
\end{equation}
\end{subequations}
where the effective packing fraction is defined by
\beq
\frac{\phi}{1-\phi}=\lambda\frac{\phi_{\eff}}{1-\phi_{\eff}}.
\label{1}
\eeq
In Eqs.~\eqref{HFEsp-Zsp} and \eqref{1}, the scaling parameters $\mu$ and $\lambda$ are
\beq
\label{mu}
\mu\equiv\frac{M_1^3M_3}{M_2^3},\quad \lambda\equiv\frac{M_1M_3}{M_2^2},
\eeq
the moments $M_n$ being defined as $M_n\equiv\sum_i x_i\sigma_i^n$.

Note that the ratio ${\phi}/(1-\phi)$ represents a \emph{rescaled} packing
fraction, namely it is the ratio between the fraction of volume, $\phi$, occupied by the spheres and the fraction of void volume, $1-\phi$. Further, $\phi [Z(\phi)-1/(1-\phi)]$ represents a (reduced) modified excess pressure with respect to a modified ideal-gas value corresponding to the fraction of void volume $1-\phi$. We refer to it as the \emph{surplus} pressure, the nomenclature having been introduced to avoid confusion with the usual excess pressure.

It is important to point out that, as discussed in Refs.~\cite{S16,SYHO17}, the \emph{surplus} approach may be generalized to deal with any $d$-dimensional HS system with dimensionality $d\neq  3$. In
this instance, one may still use Eqs.~\eqref{HFEsp-Zsp} and \eqref{1} with the packing fraction $\phi=v_d M_d$, where $v_d\equiv (\frac{\pi}{4})^{d/2}/\Gamma\left(1+\frac{d}{2}\right)$ is the volume of a $d$-sphere of unit diameter. The parameters $\mu$ and $\lambda$  are determined
by imposing consistency with the second and third virial coefficients of the mixture, leading to
\begin{equation}
\mu=\lambda^2\frac{\bar{B}_2-1}{b_2-1},\quad \lambda=\frac{\bar{B}_2-1}{b_2-1}\frac{b_3-2b_2+1}{\bar{B}_3-2\bar{B}_2+1},
\label{26}
\end{equation}
where $\bar{B}_n\equiv B_n/(v_d M_d)^{n-1}$ and $b_n\equiv B_n/(v_d \sigma^d)^{n-1}$ are reduced virial coefficients of the mixture and the monocomponent fluid, respectively ($B_n$ being the standard virial coefficients).
The approach has been applied to $d=2$ \cite{SYHO17} and $d=4,5$ \cite{HSY20} with satisfactory results. Nevertheless we will restrict ourselves to three-dimensional systems in this paper.

In the particular case of a ternary system with sizes $\sigma_1\leq\sigma_2\leq\sigma_3$, one has
\beq
\label{5}
\lambda
=\frac{(x_1\alpha_1+x_2\alpha_2+x_3)(x_1\alpha_1^3+x_2\alpha_2^3+x_3)}{(x_1\alpha_1^2+x_2\alpha_2^2+x_3)^2}\leq\frac{(1+\alpha_1)^2}{4\alpha_1},
\eeq
where $\alpha_1\equiv\sigma_1/\sigma_3$ and  $\alpha_2\equiv\sigma_2/\sigma_3$. The upper bound in Eq.~\eqref{5} corresponds to the limits  $x_1\to 1/(1+\alpha_1^2)$, $x_2\to 0$.

Now we return to our main subject. If Eq.\ \eqref{Zsp} is extended to the metastable fluid region and \emph{extrapolated} to the jamming point, where the compressibility factor diverges, one has that the random close packing  fraction ($\phi_{\text{rcp}}$) of the monocomponent system and the jamming packing fraction ($\phi_J$) of the multicomponent system  are (approximately) related by Eq.~\eqref{1}, i.e.,
\beq
 \frac{\phi_J}{1-\phi_J}\approx\lambda\frac{\phi_{\text{rcp}}}{1-\phi_{\text{rcp}}},
\label{2a}
\eeq
with $\phi_{\text{rcp}}\simeq 0.644$ \cite{BCPZ09,TS10,Z22}.
The simple ansatz \eqref{2a} allows for  a weak and a strong interpretation. According to the weak interpretation, those mixtures sharing the same value of $\lambda$ would have (approximately) the same values of $\phi_J$; the strong interpretation states that the scaled packing fraction $\phi_J/(1-\phi_J)$  is a \emph{linear} function of $\lambda$.

Of course, Eq.~\eqref{2a}  does not account for more sophisticated effects, such as the existence of rattlers, which can have a dramatic effect on the jammed packing fractions \cite{HST13}. Notwithstanding this, the  simple ansatz \eqref{2a} was found to provide an overall reasonable account of the scatter of empirical values of $\phi_J$ for discrete and continuous polydisperse mixtures \cite{SYHOO14,SYH20}.

In what concerns the \emph{densest} structures in $\ell$-component mixtures with a given set of size ratios $\{\alpha_i; i=1,\ldots ,\ell-1\}$, they are identified by the number of spheres $\{n_i; i=1,\ldots ,\ell\}$ in the unit cell. This implies that the densest packing fraction $\phi_{\max}(\{\alpha_i\},\{n_i\})$ changes with $2\ell-1$ parameters.
A number of  $\{n_1,n_2\}$ structures for the (single-phase) DBSP ($\ell=2$) at several size-ratio values $\alpha_1$ have been reported in Refs.~\cite{HJST11,HST12,KKFO21}. This has been recently complemented by $(n_1,n_2,n_3)$ structures for the (single-phase) DTSP ($\ell=3$) at several size-ratio pairs $\{\alpha_1,\alpha_2\}$ \cite{KO21,KO22}. For each structure found, the associated densest packing fraction  $\phi_{\max}$ has been obtained.
In the case of the DBSP, the values of $\phi_{\max}$ change with no clear pattern as the three parameters $\alpha_1$ and $\{n_1,n_2\}$ change. The situation is of course much more involved in the case of the DTSP, since now $\phi_{\max}$ changes with five parameters: $\{\alpha_1,\alpha_2\}$ and $\{n_1,n_2,n_3\}$.

It then seems interesting to explore the possibility that the parameter $\lambda$, as defined by Eq.~\eqref{mu} with the replacement $x_i\to n_i$, becomes useful in this context and assess to what extent $\phi_{\max}(\{\alpha_i\},\{n_i\})$ is roughly a function of the set of size ratios $\{\alpha_i\}$ and the set of numbers $\{n_i\}$ through this single parameter. According to this ansatz,
\beq
 \frac{\phi_{\max}}{1-\phi_{\max}}\approx\lambda\frac{\phi_{\text{ccp}}}{1-\phi_{\text{ccp}}},\quad \phi_{\text{ccp}}=\frac{\pi}{6}\sqrt{2}\simeq 0.7405,
\label{2b}
\eeq
in analogy with Eqs.~\eqref{1} and \eqref{2a}. The results of this exploration are presented in the following section.

\section{Results}
\label{sec2}

\begin{figure}[h!]
\includegraphics[width=0.5\textwidth]{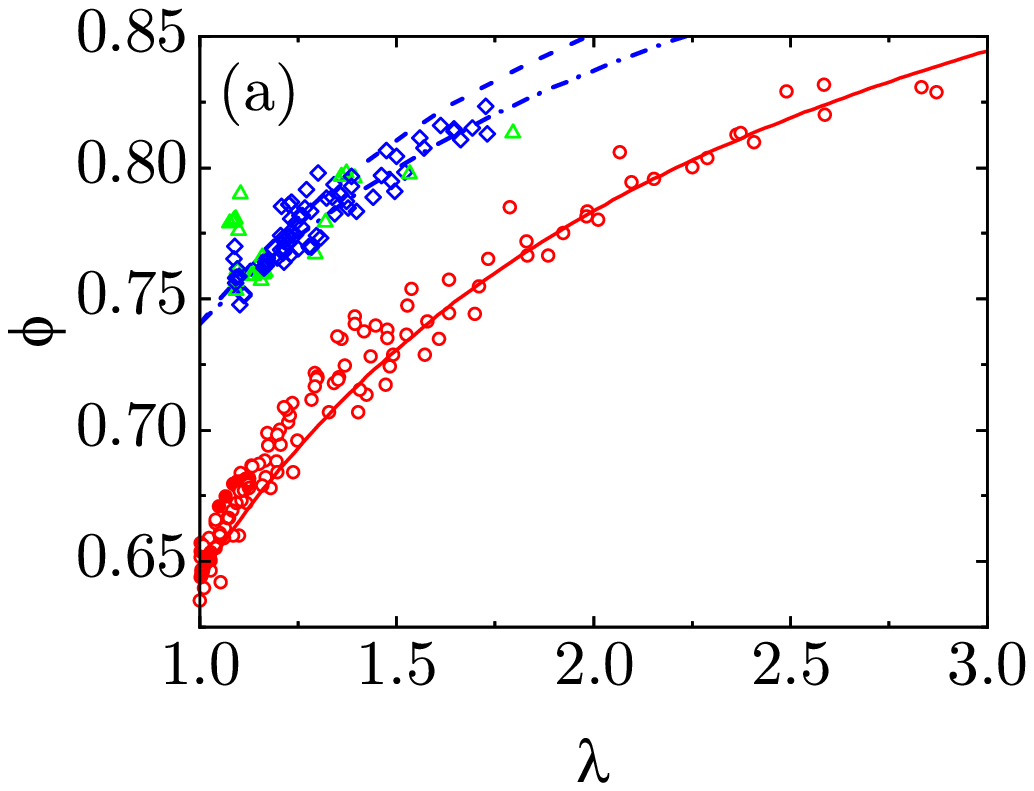}
\includegraphics[width=0.47\textwidth]{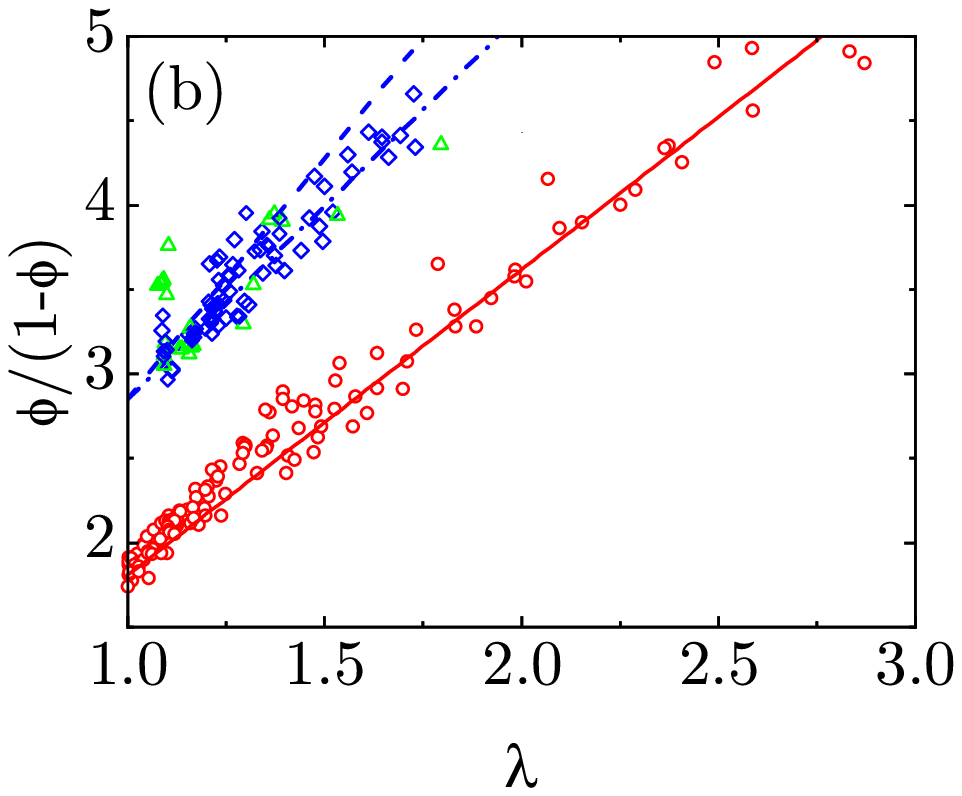}
\caption{Plot of (a) $\phi$  and (b) $\phi/(1-\phi)$  as  functions of $\lambda$ [cf.\ Eq.~\eqref{mu}]. The (red) circles are simulation results for the jamming packing fraction $\phi_J$ of polydisperse mixtures \cite{SYHOO14,SYH20}, while (green)  triangles and  (blue)  diamonds  correspond to simulation results for the densest packing fraction $\phi_{\max}$ of binary and ternary and mixtures, respectively, as reported in Refs.\ \cite{KO21,KO22}. In each panel, the (red) solid line represents the  ansatz \eqref{2a}, the (blue) dashed line represents the  ansatz \eqref{2b},  and the (blue) dash-dotted line represents the  modified ansatz \eqref{4} with $b=\frac{4}{5}$.
\label{fig1}}
\end{figure}

Figure \ref{fig1}(a) shows the values of $\phi_J$ (diverse polydisperse mixtures) and $\phi_{\max}$ (DTSP and DBST) versus the parameter $\lambda$. As expected, the ansatzes \eqref{2a} and \eqref{2b} are not strictly satisfied. However, it is certainly true that the single parameter $\lambda$  provides a useful ordering criterion for both the random and the densest packing fraction, as predicted by the weak interpretation described below Eq.~\eqref{2a}. In fact, the degree of scatter observed for $\phi_{\max}$ is comparable with that already known for $\phi_J$ \cite{SYHOO14,SYH20}. Moreover, the results displayed in Fig.\ \ref{fig1}(b) show that the scaled quantities $\phi_J/(1-\phi_J)$ and $\phi_{\max}/(1-\phi_{\max})$ present an almost linear dependence on $\lambda$ (strong interpretation).
On the other hand, it is also clear that Eq.\ \eqref{2b} tends to overestimate the  values of $\phi_{\max}$, a better performance being observed by the modified relationship
\beq
\label{4}
\frac{\phi_{\max}}{1-\phi_{\max}}\approx[1+b(\lambda-1)]\frac{\phi_{\text{ccp}}}{1-\phi_{\text{ccp}}},
\eeq
with the choice $b=\frac{4}{5}$. Note that the form of Eq.~\eqref{4} comes naturally from the requirement that
the bracketed quantity must become $1$ in the limit $\lambda\to 1$, and that such equation reduces to
Eq.~\eqref{2b} if $b = 1$.

It is also worth noticing that the agreement with Eqs.\ \eqref{2b} or \eqref{4} for the DBSP is generally worse than for the DTSP, especially if $\lambda\simeq 1.1$. Actually, it can be observed that the performance of Eqs.\ \eqref{2b} and \eqref{4} clearly tends to improve as $\lambda$ increases, that is, as the mixture deviates more from the monocomponent system.

Given the fact that the degree of scatter in the $\lambda$-representation is smaller in the ternary case than in the binary one, we can conjecture that the usefulness of the parameter $\lambda$  increases as the number of components in the alloys increases. Moreover, this conjecture relies on the fact that, as said before, in an $\ell$-component system, $\phi_{\max}$ depends on $2\ell-1$ parameters, all of them being encapsulated in the single parameter $\lambda$.
\section{Concluding remarks}
\label{sec3}

The results of the previous section deserve some further comments. One should point out that  the approximate formulae in Eqs.\ (\ref{2b}) and (\ref{4}) rest on the choice of the reference
densest monodisperse packing, which we have chosen to be the fcc crystalline close-packing value $\phi_{\text{ccp}}$. Since it is known that the densest
monodisperse packing fraction corresponds to an infinitely degenerate set of structures, namely fcc and its infinite set of stacking variants, such a choice is  problematic; fcc is a Bravais lattice, but  the densest packings are no longer Bravais lattices as one increases polydispersity from monodispersity.
Thus, there exist much more variation in the densest packing fractions than  suggested by the ansatz \eqref{2b} [or its extension, Eq.~\eqref{4}]. Nevertheless, given the heuristic character of our approach,  it turns out to be both simple enough and rooted in the values of common structures.

It is tempting to conjecture that Eqs.~\eqref{2a} and \eqref{2b}, supplemented with  $\phi_{\text{rcp}}$ for $d=2$ \cite{B83,AST14,B21,Z22} and   $\phi_{\text{ccp}}=\sqrt{3}\pi/6$, respectively, might also be useful as ordering criteria in the case of hard disks [with $\lambda$  given by Eq.~\eqref{26} with $d=2$]. While it seems worthwhile investigating this issue in the future, it lies beyond the scope of the present work.

Therefore, in conclusion, one may state that, given the simplicity of the \emph{surplus} approach, its application far outside the density region in which it was originally introduced provides a fair ordering of the available data and may serve to identify and look for geometric structures that have not been reported up to now.

\section*{Funding}
A.S. acknowledges financial support from Grant No.~PID2020-112936GB-I00 funded by MCIN/AEI/ 10.13039/501100011033, and from Grants No.~IB20079 and No.~GR21014 funded by Junta de Extremadura (Spain) and by ERDF
``A way of  making Europe.''

\section*{CRediT authorship contribution statement}
\textbf{Andr\'es Santos:} Conceptualization, Methodology, Writing - Review \& Editing, Visualization. \textbf{Mariano L\'opez de Haro:} Methodology, Validation, Writing - Original Draft.

\section*{Declaration of competing interest}
The authors declare that they have no known competing financial interests or personal relationships that could have
appeared to influence the work reported in this paper.

\section*{Data availability}
Data will be made available on request.

\section*{Acknowledgments}
We want to acknowledge a fruitful exchange of correspondence with Profs.~Salvatore Torquato and Ryotaro Koshogi which contributed to clarify some aspects of the geometric approach. We are further indebted to Prof.~Torquato for providing useful suggestions. Thanks are also due to our friend and colleague Santos Bravo Yuste for his  helpful criticisms on the contents of this paper.



  \bibliographystyle{elsarticle-num}

\end{document}